# The impedance-matched reduced acoustic cloaking with realizable mass and its layered design

Huan-Yang Chen, Tao Yang, Xu-Dong Luo, and Hong-Ru Ma

Department of Physics, Shanghai Jiao Tong University, Shanghai 200240, People's Republic of China

## Abstract

The authors present an impedance-matched reduced version of acoustic cloaking whose mass is in a reasonable range. A layered cloak design with isotropic material is also proposed for the reduced cloak. Numerical calculations from the transfer matrix methods show that the present layered cloak can reduce the scattering of an air cylinder substantially.

*To whom correspondence should be addressed. E-mail: moroshine@hotmail.com

The studies of transformation media and cloaking are very hot topics in recent years. Pendry *et al.* [1] and Leonhardt [2] first proposed the concept of transformation media in the context of electromagnetic (EM) waves independently based on the invariance of Maxwell's equations under coordinate transformation. With this concept, a scheme of achieving the EM cloaking was suggested [1] from the singular transformations given by Greenleaf *et al.* [3]. Later on, the two dimensional (2D) acoustic cloaking [4] was proposed from the one to one mapping of the Maxwell's equations and the acoustic wave equation in two dimension. This mapping has the hint that the acoustic wave equation may also be invariant under coordinate transformation [5]. Eventually, Chen and Chan [6] proved the invariance of the acoustic wave equation by mapping it to the direct current conductivity equation [3, 5]. With this mapping, the concept of acoustic transformation media was naturally obtained, which means that the acoustic waves are also freely controllable. The three dimensional (3D) acoustic cloaking was then proposed as an example of the general result. Cummer *et al.* [7] also obtained the same parameters of 3D acoustic cloaking from the Mie scattering models. However, both 2D and 3D ideal acoustic cloaking have a serious drawback that infinite mass is resulted [8, 9]. To overcome this drawback, Norris [8] suggested a kind of imperfect acoustic cloaking using finite mass by introducing small perturbations to the inner radius which are similar to the case of imperfect EM cloaking [10]. He also suggested another method [9] to realize the perfect acoustic cloaking with finite mass by using anisotropic material called "pentamode material" [5] with anisotropic stiffness.

In this letter, we propose a reduced version of acoustic cloaking by keeping its principal refractive indexes unchanged [11, 12] while the impedance of the outer boundary matches with the impedance of the background [13, 14]. This kind of reduced acoustic cloaking shows good cloaking effect and has realizable mass. An explicit design based on this reduced acoustic cloaking is also given by employing the methods of acoustic layered systems [15, 16]. The properties of the layered reduced acoustic cloaking are also studied by using the transfer matrix methods.

The parameters of the 2D ideal acoustic cloaking proposed by Cummer and Schurig [4] are,

$$\frac{\rho_r}{\rho_0} = \frac{r}{r-a}, \quad \frac{\rho_\perp}{\rho_0} = \frac{\rho_\theta}{\rho_0} = \frac{r-a}{r}, \quad \frac{\kappa_0}{\kappa} = (\frac{b}{b-a})^2 \frac{r-a}{r}, \quad (1)$$

where $\rho_0$ and $\kappa_0$ are the density and bulk modulus of the background medium, $a$ and $b$ are the inner and outer radii of the cloak. The parameters for 3D acoustic cloaking were given by Chen and Chan [6] and Cummer $et\ al.$ [7] independently as,

$$\frac{\rho_r}{\rho_0} = \frac{b-a}{b}(\frac{r}{r-a})^2, \quad \frac{\rho_\perp}{\rho_0} = \frac{\rho_\theta}{\rho_0} = \frac{\rho_\varphi}{\rho_0} = \frac{b-a}{b}, \quad \frac{\kappa_0}{\kappa} = (\frac{b}{b-a})^3 (\frac{r-a}{r})^2. \quad (2)$$

In both cases the density diverges at the inner boundary which is hard to realize in practice and reduced forms are necessary in order to manufacture the acoustic cloaking. The reduction from ideal cloaking is to loosen the requirements on density and modulus by combining these two requirements into one through the principal refractive index. The principal refractive indexes of acoustic waves, defined by the ratio of density and bulk modulus, can be written in a unified form for both 2D and 3D acoustic cloak as

$$\frac{n_r^2}{n_0^2} = \frac{\rho_\perp/\kappa}{\rho_0/\kappa_0} = (\frac{b}{b-a})^2 (\frac{r-a}{r})^2, \quad \frac{n_\perp^2}{n_0^2} = \frac{\rho_r/\kappa}{\rho_0/\kappa_0} = (\frac{b}{b-a})^2. \quad (3)$$

This formula forms the key starting point of our proposed reduced acoustic cloaking. By loosening the requirements on the density and modulus independently, the cloaking is no longer ideal. However, we may use this freedom to design acoustic cloaking which has some imperfections but is realizable in practice. Suppose that $\kappa$ is a continuous function of $r$, i.e. $\frac{\kappa}{\kappa_0} = \lambda(r)$, to obtain,

$$\frac{\rho_r}{\rho_0} = \lambda(r)(\frac{b}{b-a})^2, \quad \frac{\rho_\perp}{\rho_0} = \lambda(r)(\frac{b}{b-a})^2(\frac{r-a}{r})^2, \qquad (4)$$

for both 2D and 3D acoustic cloaking. The cloaks with the above parameters are called the reduced cloak. The guide line for choosing a proper $\lambda(r)$ is to match the impedance on outer boundary between the cloak and the background media. We can simply choose $\frac{\kappa}{\kappa_0} = \lambda(r) = \frac{b-a}{b}$ to realize an impedance-matched version. With this choice, the parameters become,

$$\frac{\rho_r}{\rho_0} = \frac{b}{b-a}, \quad \frac{\rho_\perp}{\rho_0} = \frac{b}{b-a}(\frac{r-a}{r})^2, \quad \frac{\kappa}{\kappa_0} = \frac{b-a}{b}. \qquad (5)$$

This set of parameters requires only $\rho_\perp$ to be position dependent. The density tensor is finite for all positions and realizable for the future designs. It should be noted that the above version is not the only design of an impedance-matched reduced cloak. Other choices of $\kappa$ is also possible to get the impedance-matched design.

Now we come to design a feasible acoustic cloak using acoustic layered systems. The acoustic layered systems with two kinds of alternating isotropic materials (A and B) can produce a transversely isotropic system whose effective density becomes a tensor. Suppose the densities of the isotropic materials are $\rho_A$ and $\rho_B$, and the bulk modulus are $\kappa_A$ and $\kappa_B$. Then a structure consists of concentric layers of A and B with the same thicknesses, which are much smaller compared to the incident wavelength, has the following effective density tensor and bulk modulus [15, 16],

$$\rho_r = \frac{\rho_A + \rho_B}{2}, \quad \rho_\perp = \frac{2\rho_A \rho_B}{\rho_A + \rho_B}, \quad \kappa = \frac{2\kappa_A \kappa_B}{\kappa_A + \kappa_B}. \tag{6}$$

The design of the reduced cloak is a structure of concentric multilayer of the effective AB layers with different effective material parameters. Using the two-step approach [15, 17], we can obtain the required density and modulus for each layer of the cloak. Suppose the cloak consist 2N layers of isotropic materials. Then the densities of each layer are,

$$\frac{\rho_{2i-1}}{\rho_0} = \frac{b}{b-a}(1 - \sqrt{1 - (\frac{r_{2i-1} - a}{r_{2i-1}})^2}), \quad i=1,2,3,...,N, \tag{7a}$$

$$\frac{\rho_{2i}}{\rho_0} = \frac{b}{b-a}(1 + \sqrt{1 - (\frac{r_{2i} - a}{r_{2i}})^2}), \quad i=1,2,3,...,N, \tag{7b}$$

where $r_i = a + (2i-1)\frac{b-a}{4N}$, $i = 1,2,3,...,2N$. All the layers share the same modulus $\frac{\kappa}{\kappa_0} = \frac{b-a}{b}$. Notice that for the reduced cloak the density ratios $\rho_i / \rho_0$ should locate at the range [0, $\frac{b}{b-a}(1 - \sqrt{1 - (\frac{b-a}{b})^2})$] and [$\frac{b}{b-a}(1 + \sqrt{1 - (\frac{b-a}{b})^2})$, $\frac{2b}{b-a}$]. To be concrete and comparable with results of ideal case, we now focus on the 2D case and use parameters in ref [15]. The number of effective AB layers is N=20 and b=2a=2m. The density ratios locate at [0, 0.27] and [3.74, 4], which are more feasible than the ideal case whose density ratios locate at [0.025, 0.267] and [3.73, 39.99]. The maximum of the density in the reduced case is much smaller than the one in the ideal case since the infinite mass no longer exists in the reduced case. The plane wave is incident from left to right, with a wavelength equals to 1 m. The background is water, $\rho_0 = 998 \ kg/m^3$ and $\kappa_0 = 2.19 \ GPa$. Since the air bubble will cause great scattering in the water, we will use the present layered cloak to reduce the scattering of an air core and show the good cloaking effect. The parameters of air are, $\rho_{air} = 1.25 \ kg/m^3$ and $c_{air} = 343 \ m/s$.

Fig. 1a shows the calculated acoustic pressure field distribution for incident plane wave scattered by an air cylinder with the radius a=1m in the water. From the figure, we see that the scattering in this case is very large. Fig. 1b shows the acoustic pressure field distribution for incident plane wave scattered by the air cylinder surrounded with the present layered cloak. The scattering is greatly reduced when compared to Fig. 1a.

The scattering amplitude $f(\theta)$ [18] normalized by that of the incident wave are plotted in Fig. 2. The dashed black line shows the one for a bare air cylinder in water. It causes large scattering for all the angles especially for the forward direction. The solid red line denotes the scattering amplitude for the scaterer surrounded with the layered cloak from the ideal case [15]. And it reduces the scattering a lot for all the angles. However, there are still some detectable scattering in the forward direction. That is due to the small number of layers, it will be improved if the layer number is increased. The blue dotted line denotes the scattering amplitude for the scaterer surrounded with the present layered cloak. It also has good cloaking effect for all the angles except for some detectable scattering in the forward direction. Part of this scattering is due to the small number of layer as in the ideal case, and another contribution is from the reduced version intrinsically as will be discussed below.

The total cross section $\sigma$ [18] of the bare scaterer is 0.7255, when the scaterer is surrounded with the layered cloak from the ideal case, its total cross section is reduced to 0.035 (which is 4.82% of the bare case), when the scaterer is surrounded with the present layered cloak, its total cross section becomes 0.0849 (which is 11.7% of the bare case). Fig. 3a shows the scattering amplitude $f(\theta=0)$ for cloaks of different layer number 2N, which denotes the scattering in the forward direction. When 2N is large enough, the scattering amplitude $f(\theta=0)$ for ideal case approaches to zero while the one for reduced case tends to a very small positive value which is about 2. This is the intrinsic value due to the reduction of cloaking from the ideal case. The relationship between the total scattering cross section and 2N is similar with the forward scattering amplitude, which is plotted in Fig. 3b. All the above calculations are from the well-known transfer matrix methods [18]. Extension of the current reduced acoustic cloaking and the layered system design to 3D is straight forward, and the results are similar.

In conclusion, we have proposed an impedance-matched reduced acoustic cloaking with realizable mass which will be useful for the future design. We suggested a design using acoustic layered systems and studied the layered cloak using the transfer matrix methods. The present layered cloak shows good cloaking effect.

## Acknowledgements

This work was supported by the National Natural Science Foundation of China under grand No.10334020 and in part by the National Minister of Education Program for Changjiang Scholars and Innovative Research Team in University.

## Reference

[1] J. B. Pendry, D. Schurig, and D. R. Smith, Science **312**, 1780 (2006).

[2] U. Leonhardt, Science **312**, 1777 (2006).

[3] A. Greenleaf, M. Lassas, and G. Uhlmann, Math. Res. Lett. **10**, 685 (2003); A. Greenleaf, M. Lassas, and G. Uhlmann, Physiol. Meas **24**, 413 (2003).

[4] S. A. Cummer and D. Schurig, New J. Phys. **9**, 45 (2007).

[5] G. W. Milton, M. Briane, and J. R. Willis, New J. Phys. **8**, 248 (2006).

[6] H. Y. Chen and C. T. Chan, Appl. Phys. Lett. **91**, 183518 (2007).

[7] S. A. Cummer, B.-I. Popa, D. Schurig, D. R. Smith, J. B. Pendry, M. Rahm, and A. Starr, Phys. Rev. Lett. **100**, 024301 (2008).

[8] A. N. Norris, URL http://arxiv.org/abs/0802.0701 (2008).

[9] A. N. Norris, URL http://arxiv.org/abs/0805.0080 (2008), published online on Proc. R. Soc. A 2008.

[10] H. Y. Chen, Z. Liang, P. Yao, X. Jiang, H. R. Ma and C. T. Chan, Phys. Rev. B **76**, 241104(R) (2007).

[11] S. A. Cummer, B.-I. Popa, D. Schurig, D. R. Smith, and J. B. Pendry, Phys. Rev. E 74, 036621 (2006); D. Schurig, J. J. Mock, B. J. Justice, S. A. Cummer, J.

B. Pendry, A. F. Starr, and D. R. Smith, Science **314**, 977 (2006).

[12] W. Cai, U. K. Chettiar, A. V. Kildishev, and V. M. Shalaev, Nat. Photonics **1**, 224 (2007).

[13] W. Cai, U. K. Chettiar, A. V. Kildishev, V. M. Shalaev, and G. W. Milton, Appl. Phys. Lett. **91**, 111105 (2007).

[14] M. Yan, Z. Ruan, and M. Qiu, Opt. Express **15**, 17772 (2007).

[15] Y. Cheng, F. Yang, J. Y. Xu, and X. J. Liu, Appl. Phys. Lett. **92**, 151913 (2007).

[16] M. Schoenberg and P. N. Sen, J. Acoust. Soc. Am. **73**, 61 (1983).

[17] Y. Huang, Y. Feng and T. Jiang, Opt. Express **18**, 11133 (2007).

[18] L.-W. Cai and J. Sánchez-Dehesa, New J. Phys. **9**, 450 (2007).

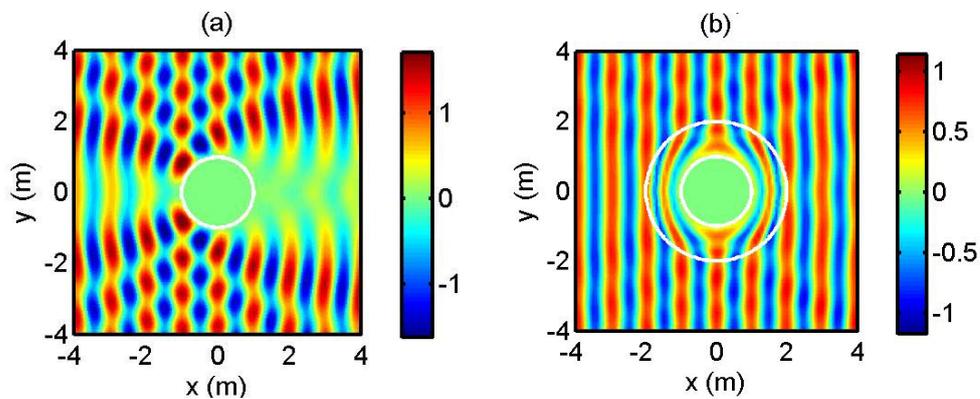

Fig. 1 (Color online) Acoustic pressure field distribution near an air cylinder (a) without cloak; (b) with the present layered cloak.

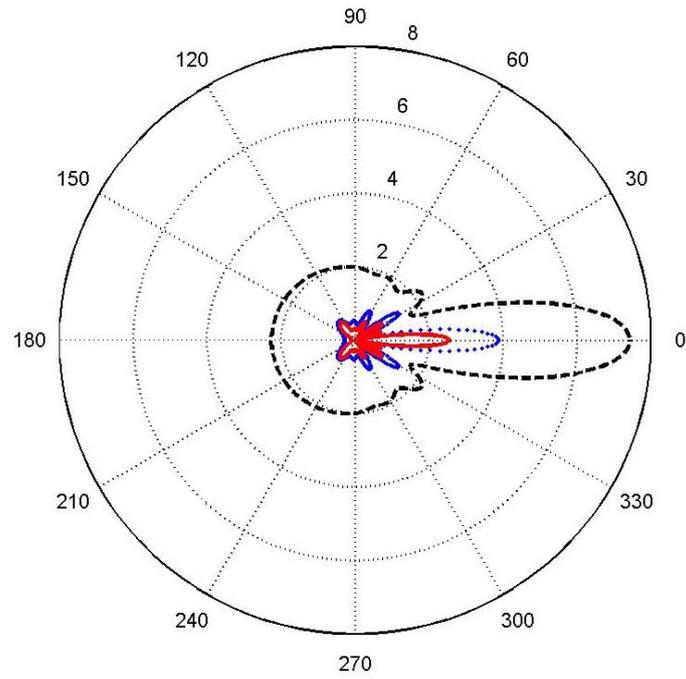

Fig. 2 (Color online) Scattering form factors for three cases: bare air cylinder only (denoted by dashed black line), the scaterer with layered cloak from ideal case (denoted by solid red line), the scaterer with the present layered cloak (denoted by dotted blue line).

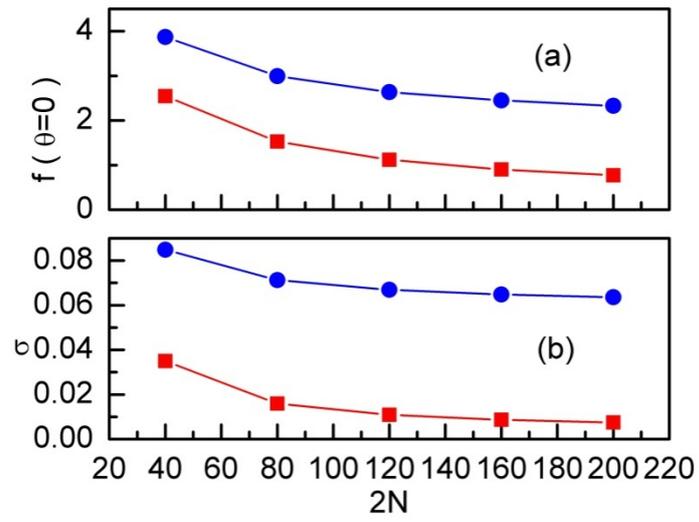

Fig. 3 (Color online) (a) Scattering form factors for the forward direction for cloaks with different layer number (2N). (b) Total scattering cross section for cloaks with different layer number (2N). The blue circle symbols denote the reduced case while the red squire symbols denote the ideal case.